\documentclass[lettersize,journal]{IEEEtran}

\usepackage{amsmath,amsfonts}
\usepackage{algorithmic}
\usepackage{algorithm}
\usepackage{array}

\usepackage{textcomp}
\usepackage{stfloats}
\usepackage{graphicx}
\usepackage{subcaption}
\usepackage[justification=centering]{caption}
\usepackage[style=ieee,sorting=none,doi=false,url=false]{biblatex}

\usepackage{hyperref}

\usepackage{url}
\usepackage{verbatim}

\begin{document}

\title{Design of an $M$-ary Chaos Shift Keying System Using Combined Chaotic Systems}

\author{Tingting Huang, Jundong Chen, Huanqiang~Zeng,~\IEEEmembership{Senior~Member,~IEEE, }\\Guofa~Cai,~\IEEEmembership{Senior~Member,~IEEE, }and Haoyu Zhou
\thanks{Corresponding author: Huanqiang Zeng.}
\thanks{Tingting Huang, Jundong Chen, Huanqiang Zeng and Haoyu Zhou are with the School of Engineering, Huaqiao University, Quanzhou 362021, China
(e-mail: tthuang@hqu.edu.cn; JDchan@163.com; zeng0043@hqu.edu.cn; 18890584877@163.com).}
\thanks{GuoFa Cai is with the School of Information Engineering, Guangdong
University of Technology, Guangzhou 510006, China (e-mail: caiguofa2006@gdut.edu.cn).}
}



\maketitle
\begin{abstract}

In traditional chaos shift keying (CSK) communication systems, implementing chaotic synchronization techniques is costly but practically unattainable in a noisy environment. This paper proposes a combined chaotic sequences-based $M$-ary CSK (CCS-$M$-CSK) system that eliminates the need for chaotic synchronization. At the transmitter, the chaotic sequence is constructed by combining two chaotic segments of different lengths, where each is generated from distinct chaotic systems and only one kind of chaotic segment modulates the information signal. At the receiver, a deep learning unit with binary classification is meticulously designed to recover information symbols. The symbol error rate (SER) performance of the proposed system is evaluated over additive white Gaussian noise (AWGN) and multipath Rayleigh fading channels. Specifically, the impact of varying misalignment lengths on the SER performance of the system is analyzed when the received sequence is misaligned. Furthermore, the proposed system demonstrates significant performance advantages over existing CSK-based systems in multipath Rayleigh fading channels. These features establish CCS-$M$-CSK as a promising candidate for various applications, including Vehicle-to-Everything (V2X).

\end{abstract}


\section{Introduction}

\IEEEPARstart{O}{ver} the last two decades, chaos signals have gained significant attention due to their remarkable properties, including non-periodicity, wide-band nature, initial sensitivity and robustness against eavesdropping by intruders \cite{r1-7,r1-8, r1-6}. To circumvent the challenges of chaotic synchronization in CSK systems, researchers have increasingly focused on chaotic incoherent systems. Among all noncoherent schemes, differential chaos shift keying (DCSK)-based systems stand out for their simple architectures and low hardware complexity, making them appealing for practical applications such as the internet of things (IoT) \cite{r1-4}, vehicle-to-vehicle (V2V) \cite{r1-2}, simultaneous wireless information and power transfer (SWIPT) \cite{r1-3}, power line communication (PLC) \cite{r1-1}.

However, a significant drawback of the existing DCSK systems is the repeated transmissions of reference signals, which reduces of data rate, energy efficiency, reliability, and security. Current research primarily focuses on enhancing bandwidth efficiency (BE) in DCSK-based communication while achieving a reliable bit error rate (BER) with minimum energy per bit. Employing $M$-ary transmission is a promising solution to achieve the aforementioned performance. In $M$-ary DCSK systems, 2-dimensional and $M$-dimensional constellations are widely adopted. $M$-ary modulation schemes often employ Walsh code \cite{r2-5,r2-6,r2-7} or Gram-Schmidt algorithm\cite{r2-4} to construct $M$-dimensional orthogonal bases, while conventional sinusoidal signals \cite{r2-11} or Hilbert transform \cite{r2-8,r2-9,r2-10} are typically used for 2-dimensional bases. Recently, a three-dimensional constellation $M$-ary DCSK system based on the Gram-Schmidt algorithm was proposed in \cite{r2-6}. Additionally, maintaining orthogonality between the chaotic reference and the information signal is also crucial, and this is typically achieved using orthogonal codes \cite{r2-5}, time slots \cite{r2-12}, and frequency \cite{r2-10,r2-11}. For $M$-ary DCSK systems, the most important work is to maintain orthogonality.

As aforementioned, overcoming the deficiencies of DCSK systems is quite costly, and the performance loss caused by their inherent structure cannot be fully mitigated. For example, noise introduced into the reference signal inevitably degrades performance, while the presence of the reference signal compromises system security. Consequently, researchers have redirected their attention to CSK systems, seeking to overcome their technical challenges using advanced tools such as artificial intelligence (AI). A 2-ary CSK system employing orthogonal frequency-division multiplexing (OFDM) was proposed in \cite{r3-1}, where a single chaotic map is used at the transmitter, and an long short-term memory (LSTM)-based classifier recovers information data. This approach was later extended to an $M$-ary scheme in \cite{r3-2}, requiring $M$ chaotic maps to construct an $M$-dimensional signal space, where a convolutional neural network (CNN)-based classifier is used for data recovery. However, selecting a set of quasi-orthogonal chaotic maps is challenging, and as noted by the authors, the BER performance deteriorates significantly as $M$ increases.


Inspired by these ideas, we propose a combined chaotic sequences based $M$-ary CSK (CCS-$M$-CSK) system. At the transmitter, the transmitted chaotic sequence consists of two chaotic segments derived from different chaos systems, each serving distinct functions and having different lengths. The shorter segment is used to modulate the information bits, while the longer segment is employed to conceal the valid information of the short segment. Unlike the approaches in \cite{r3-1} and \cite{r3-2}, where the transmitter in \cite{r3-1} utilizes a single chaotic system and the transmitter in \cite{r3-2} employs $M$ chaotic systems, our system introduces a more flexible structure. Additionally, since the length of the long chaotic signal is significantly greater than that of the short chaotic signal, the short chaotic signal can be embedded at any position within the long chaotic signal, thereby enhancing security. As a result, depending on the position of the short chaotic signals, our system can achieve $M$-ary transmission.

The main contributions of this paper are summarized as follows:

\begin{enumerate}
\item We propose a CCS-$M$-CSK system to enhance data rate and reliability. In the proposed system, the chaotic sequence comprises two segments of different lengths, where each is generated by distinct chaotic systems. The shorter segment is embedded at variable positions within the longer chaotic segment based on a symbol-to-position mapping table, enabling efficient $M$-ary transmission.

\item We propose a novel deep neural networks (DNN)-based receiver that integrates convolutional feature extraction with a self-attention mechanism for robust chaotic sequence detection. This architecture eliminates the reliance on orthogonal basis while delivering superior detection performance across diverse channel conditions.

\item We analyze the impact of key parameters on the system's symbol error rate (SER) and evaluate its tolerance to mismatch in the length of received data. Additionally, we compare the performance of the proposed system with other CSK systems, which demonstrates that the proposed system achieves superior performance.

\end{enumerate}


\section{CCS-$M$-CSK System}


\subsection{Transmitter}

\begin{figure}[!t]
\centering
\includegraphics[width=\columnwidth]{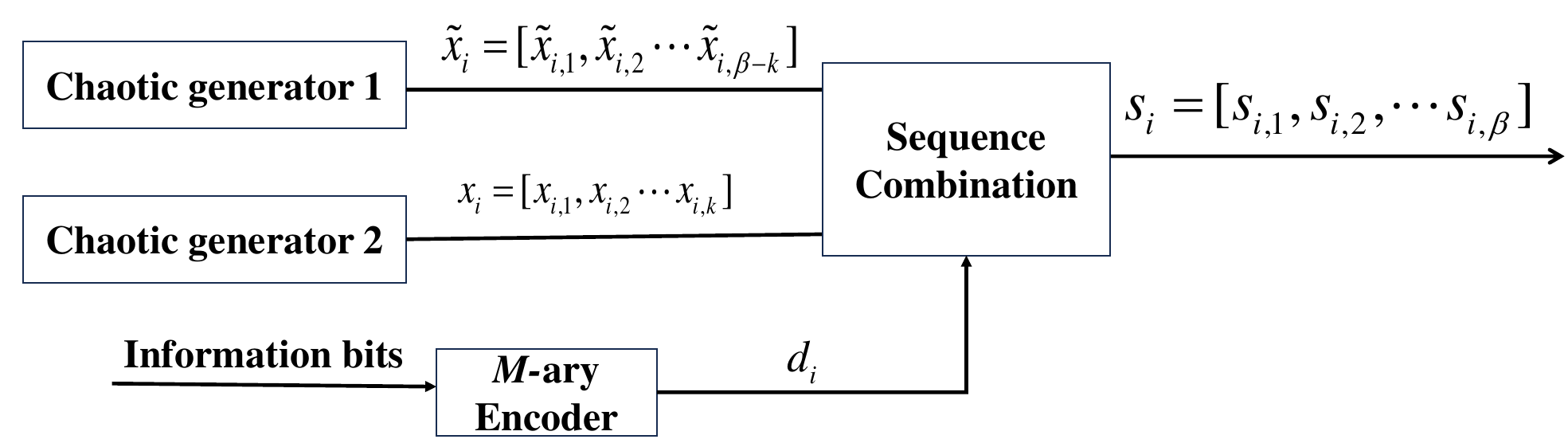}
\captionsetup{justification=raggedright,singlelinecheck=false}
\caption{The transmitter of CCS-$M$-CSK system.}
\label{fig_1}
\end{figure}

It has been proven that a sequence formed by combining multiple chaotic sequences from different chaotic systems remains chaotic. \cite{r4-1}. Based on this theory, the transmitted signal  $\boldsymbol{s}$ is a combination of a Logistic sequence $\boldsymbol{\tilde{x}}$ and a Cubic sequence $\boldsymbol{x}$. The transmitter architecture, illustrated in Fig.~\ref{fig_1}, includes Chaotic Generator 1, which produces a Logistic sequence $\boldsymbol{\tilde{x}}$ of length  $\beta-k$ and Chaotic Generator 2, which generates a Cubic sequence $\boldsymbol{x}$ of length $k$. The Logistic sequence $\boldsymbol{\tilde{x}}$ and Cubic sequence $\boldsymbol{x}$ can be indicted individually as
\begin{subequations}
\begin{equation}
\tilde{x}(n+1) = 3.7 \times \tilde{x}(n) \times (1 -\tilde{x}(n)),
\end{equation}
\begin{equation}
x(n+1) = 4 \times x(n)^3 + 3 \times x(n).
\end{equation}
\end{subequations}

\begin{figure}[!t]
\centering
\includegraphics[width=2.5in]{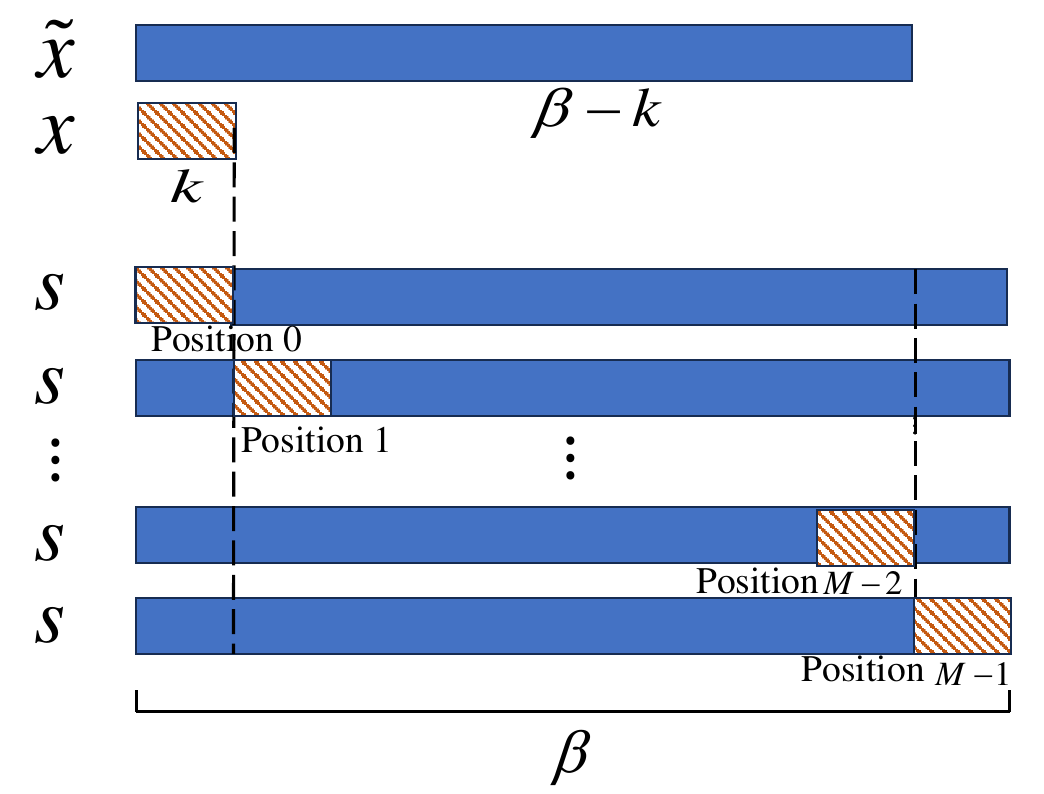}
\captionsetup{justification=raggedright,singlelinecheck=false}
\caption{Composition of transmit signals corresponding to different information symbols.}
\label{fig_2}
\end{figure}

\begin{figure}[!t]
\centering
\includegraphics[width=2in]{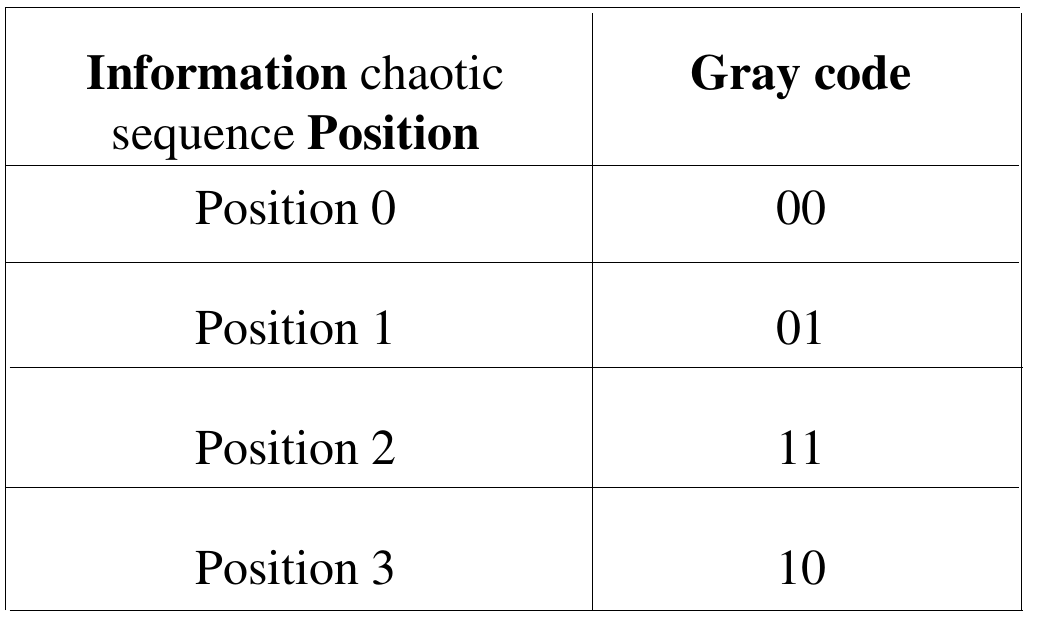}
\captionsetup{justification=raggedright,singlelinecheck=false}
\caption{Mapping table of symbols to information chaotic sequence positions when $M = 4$.}
\label{fig_3}
\end{figure}

Additionally, the $M$-ary encoder converts the bit streams into $M$-ary symbols using Gray coding, while the sequence combination unit generates a combined chaotic sequence  $\boldsymbol{s}$ based on the symbol-to-position mapping table, as shown in Fig.~\ref{fig_2} and Fig.~\ref{fig_3}.

To be more specific, the $i$-th bit stream is firstly coded by $M$-ary encoder to generate symbol $d_i$. Then the transmit signal $\boldsymbol{s}_i=[s_{i,1},s_{i,2},\cdots, s_{i,\beta}]$ matching $d_i$ is generated by sequence combination unit according to the mapping table of symbols. For simplicity in analysis, let $\beta = Mk$, where $M=2^n$ and $n\in \mathbb{N}$. $\boldsymbol{s_i}$ can be indicted as 
\begin{equation}
s_i(q) = 
\begin{cases}
x_i(q), & q = (c-1)k + j, \\
& c = 1,2,\ldots,M, \; j = 1,2,\ldots,k \\
\tilde{x}_i(q), & 1 \leq q \leq \beta, \; q \neq (c-1)k + j.
\end{cases}
\end{equation}

It should be noted that this mapping table is available for both transmitter and receiver.

\subsection{DNN-based Receiver}

The receiver architecture is illustrated in Fig.~\ref{fig_5}. When the transmitted signal $\boldsymbol s_i$ propagates through the multipath Rayleigh fading channel, the received signal $\boldsymbol r_{i}$ is expressed as:
\begin{equation}
r_i(q) = \sum_{g=1}^{G} \alpha_{i,g}\times s_i(q-\tau_{i,g}) + n_i(q), 1\leq q\leq\beta,
\end{equation}
where $\alpha_{i,g}$ and $\tau_{i,g}$ represents channel gain and delay of the $g$-th path, respectively. $G$ is the number of paths, and
$n_i(q)$ denotes additive white gaussian noise (AWGN) with zero mean and variance $N_0/2$. The receiver uses a DNN-based intelligent demodulator. The demodulator first evenly splits received sequence $r_i$ into $M$ subsequences with equal length $k$. These subsequences are then processed by a pre-trained DNN, which outputs the probability vector:
\begin{equation}
\mathbf{p} = [p_1, p_2, ..., p_M],
\end{equation}
where $p_j$ denotes the probability that the $j$-th subsequence contains the embedded short chaotic sequence $\boldsymbol{x}$. Since the chaotic information signal appears in only one position, the estimated symbol $\hat{d}_i$ is determined by
\begin{equation}
\hat{d}_i = \arg\max_{j} p_j.
\end{equation}

This deep learning-based method significantly improves the demodulation performance and anti-interference ability of the system in complex channel environments. The detailed DNN architecture and training process will be described as follows.

\subsection{DNN Hierarchical Arrangement}

\begin{figure}[!t]
\centering
\subfloat[]{\includegraphics[width=2.5in]{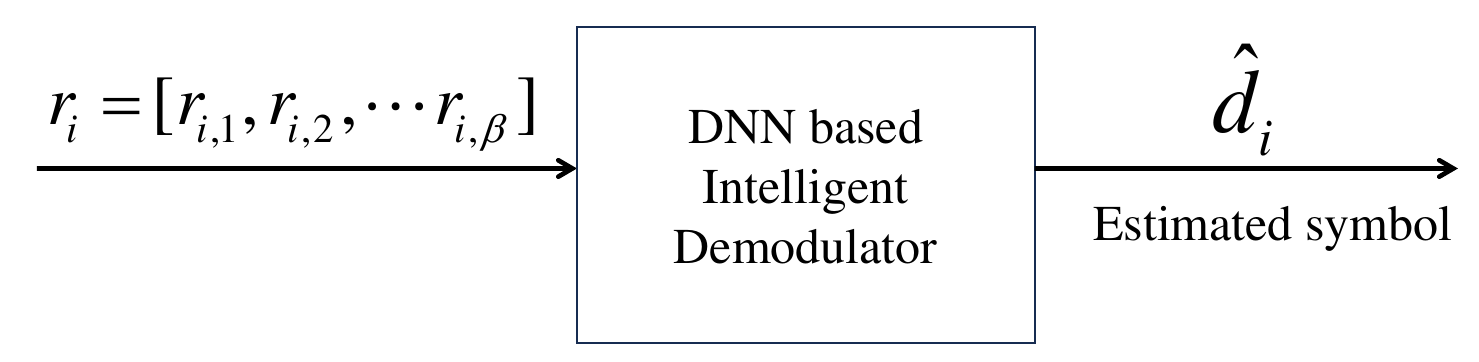}} \\
\subfloat[]{\includegraphics[width=2.6in]{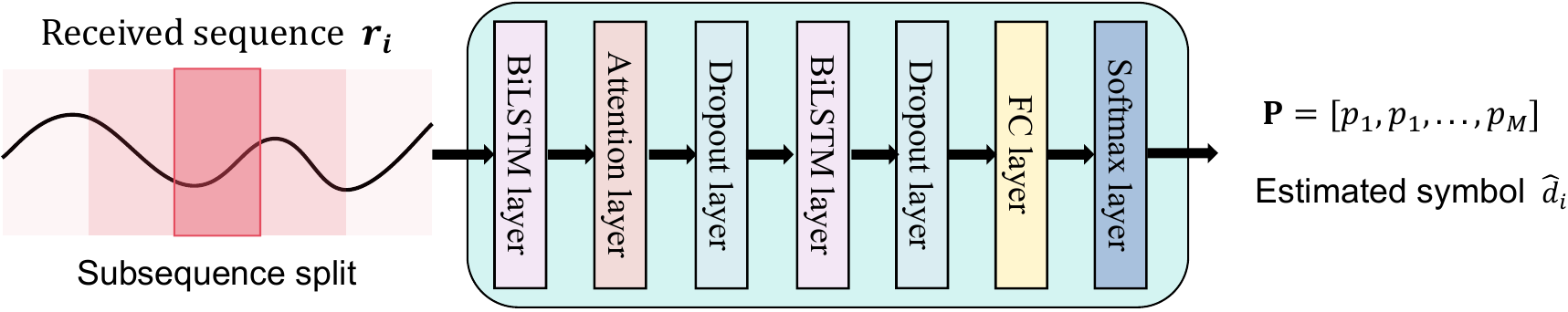}}
\captionsetup{justification=raggedright,singlelinecheck=false}
\caption{The receiver of CCS-$M$-CSK system. (a)  DNN-based receiver. (b) DNN architecture.}
\label{fig_5}
\end{figure}

The proposed DNN architecture, as illustrated in Fig. \ref{fig_5}(b), is specifically designed to detect and identify the chaos detection sequence embedded within the received signal under complex channel conditions. The network begins with a sequence input layer that splits inputs, followed by a bidirectional LSTM (BiLSTM) layer with 64 hidden units to capture temporal dependencies in both directions. This BiLSTM effectively extracts features from both past and future contexts of the chaotic sequence. A self-attention layer (64 heads, 256 dimension) is then employed to weigh significant temporal features, enabling the model to focus on the most relevant parts of the sequence for detection. Dropout layers ($p=0.2$) are strategically inserted after each major processing block for regularization, preventing co-adaptation of hidden units during training. A second BiLSTM layer with 64 hidden units processes the attended features, maintaining only the final state to compress the temporal information into a fixed-dimensional representation. The architecture concludes with a fully connected layer followed by softmax activation, producing probability $p_j$ for each subsequence.




\subsection{Offline Training}

The training process is conducted offline using simulated chaotic sequences and channel effects. The training set consists of $2\times10^5$ sequence pairs ${\boldsymbol{r}_i, d_i}$, where $\boldsymbol{r}_i$ represents the received subsequence and $d_i\in{\{0,1\}}$ denotes sequences generated by Cubic map ($d_i=1$) or Logistic map ($d_i=0$). For channel conditions, the signal-to-noise ratio (SNR) is set to [12, 14] dB under AWGN channel and [14, 16] dB under Rayleigh fading channel. During the offline training phase, the DNN parameters are updated via iterative training with the training dataset. Let $\boldsymbol{D}={\{d_1,d_2,...,d_N\}}$ and $\boldsymbol{\hat{D}}={\{\hat{d_1},\hat{d_2},...,\hat{d}_N\}}$ represent the true labels and estimated data, respectively. the network is optimized by minimizing the binary cross-entropy loss between $\boldsymbol{D}$ and $\boldsymbol{\hat{D}}$.

The network is trained using Adam optimizer with an initial learning rate of 0.001 and an early stopping mechanism. The training process uses mini-batches of size 128 and 20\% of the data is reserved for validation. All experiments are conducted using MATLAB 2023b and CUDA 12.0.

\subsection{Complexity Analysis}

\begin{table}[!t]
\centering
\caption{NETWORK ARCHITECTURE SUMMARY}
\label{table_1}
\begin{tabular}{|l|c|c|}
\hline
Layer & Tensor Shape & Activation \\
\hline
Input Layer & 2($C$) × $T$ & - \\
BiLSTM Layer 1 & 128($C$) × $T$ & -  \\
Self-Attention Layer & 128($C$) × $T$ & -  \\
Dropout Layer 1 & - & - \\
BiLSTM Layer 2 & 128($C$) × $T$ & - \\
Dropout Layer 2 & - & - \\
Fully Connected Layer & 2($C$)  & Softmax \\
\hline
\noalign{\vskip 1mm} 
\multicolumn{3}{l}{\small Note: $C$: Channels, $T$: Sequence length} \\
\end{tabular}
\label{tab:network_architecture}
\end{table}

For each input sequence of length $T$, the received vectors are first separated into real and imaginary components, resulting in two input channels. Based on \cite{r3-1}, the computational complexity of the proposed DNN architecture is calculated as $\mathcal{O}(\sum_{l=1}^{L_1}(N_{in}N_h + N_h^2)T + (QK^T)T^2 + (QK^T V)T + N_hC)$, where $L_1=2$ is the number of BiLSTM layers, $N_{in}$ is the input dimension, $N_h=128$ is the number of hidden units, $T$ is the sequence length, and $Q$, $K$, $V$ represent the dimension of attention mechanism parameters, $C=2$ is the number of output classes. When sequence length $T=128$, by substituting the hyperparameters from Table \ref{tab:network_architecture}, the complexity of the proposed architecture is approximated as $\mathcal{O}(2.1M^3)$. Although this complexity is higher than traditional non-coherent detectors, the proposed architecture can be efficiently implemented on specialized hardware like GPUs or FPGAs for practical applications.

\subsection{Security Evaluation}
We analyze the security of the proposed system by considering information leakage and brute-force eavesdropping. Assuming that each symbol is sent with equal probability, the information leakage rate $L(X;R_e)$ between the transmitted symbol $X$ and the symbol $R_e$ received by the eavesdropper can be expressed as:
\begin{equation}
\begin{aligned}
 L(X;R_e) &= H(X) - H(X|R_e) \\
         &= 1 + P_e \log2(P_e) + (1-P_e)\log2(1-P_e).
\end{aligned}
\end{equation}
where $H(.)$ denotes the entropy operation and $P_e$ denotes the eavesdropper's BER. 


\section{Results and Discussions}
This section evaluates the SER performance of the proposed CCS-$M$-CSK system under AWGN and two-path Rayleigh channels. In a two-path Rayleigh channel, with identical average power gains, is considered. The average power gain in each path is assumed to be 0.5, (i.e., $E(\alpha_1)=E(\alpha_2)=0.5$), with  $\tau_1 = 0$, and $\tau_2 = 2$. We assume that the channel coefficients ($\alpha_l$) and training SNR change after each transmission.

\subsection{Impact of Various Parameters on SER Performance}
\begin{figure}[!t]
\centering
\begin{subfigure}{0.49\columnwidth}    
    \includegraphics[width=\textwidth,height=1.6in]{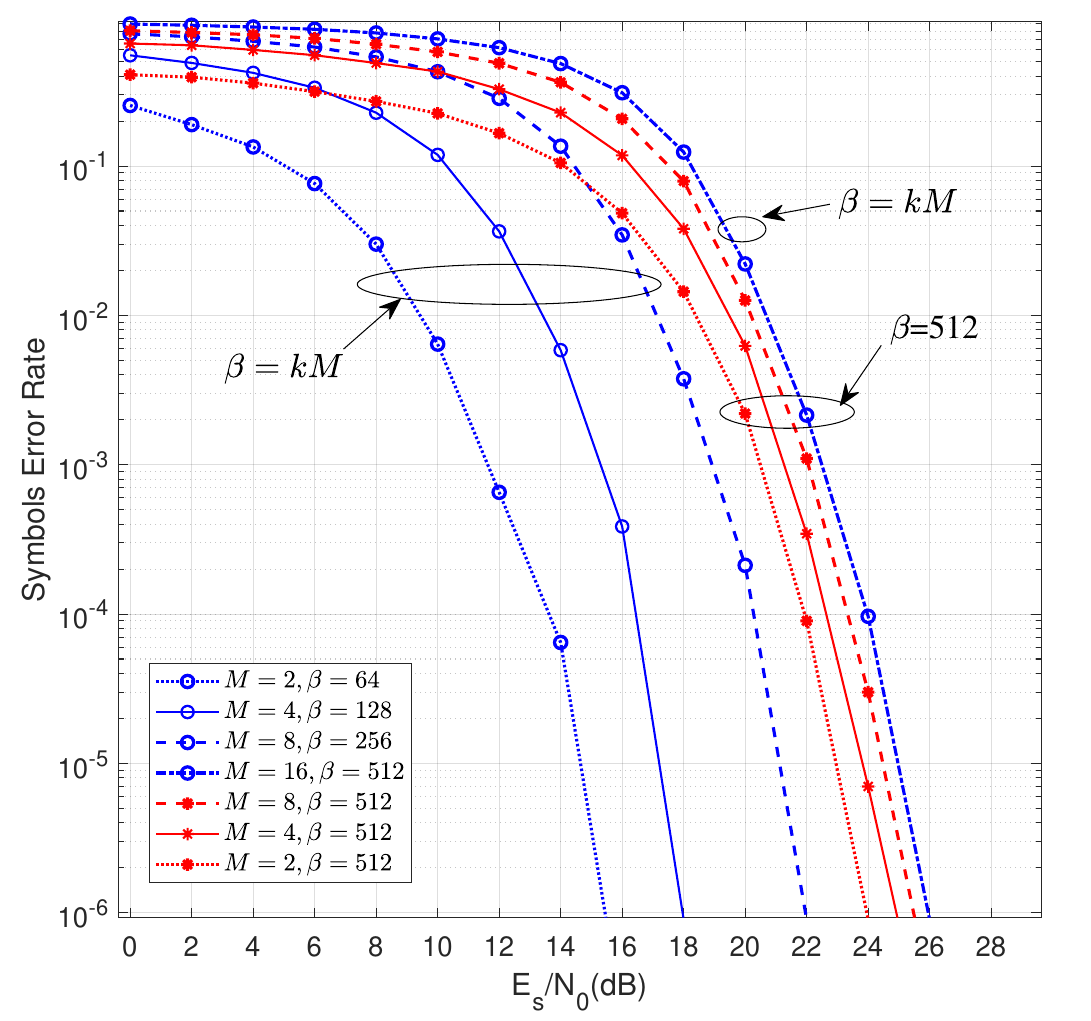}
    \caption{}
\end{subfigure}
\hfill
\begin{subfigure}{0.49\columnwidth}    
    \includegraphics[width=\textwidth,height=1.6in]{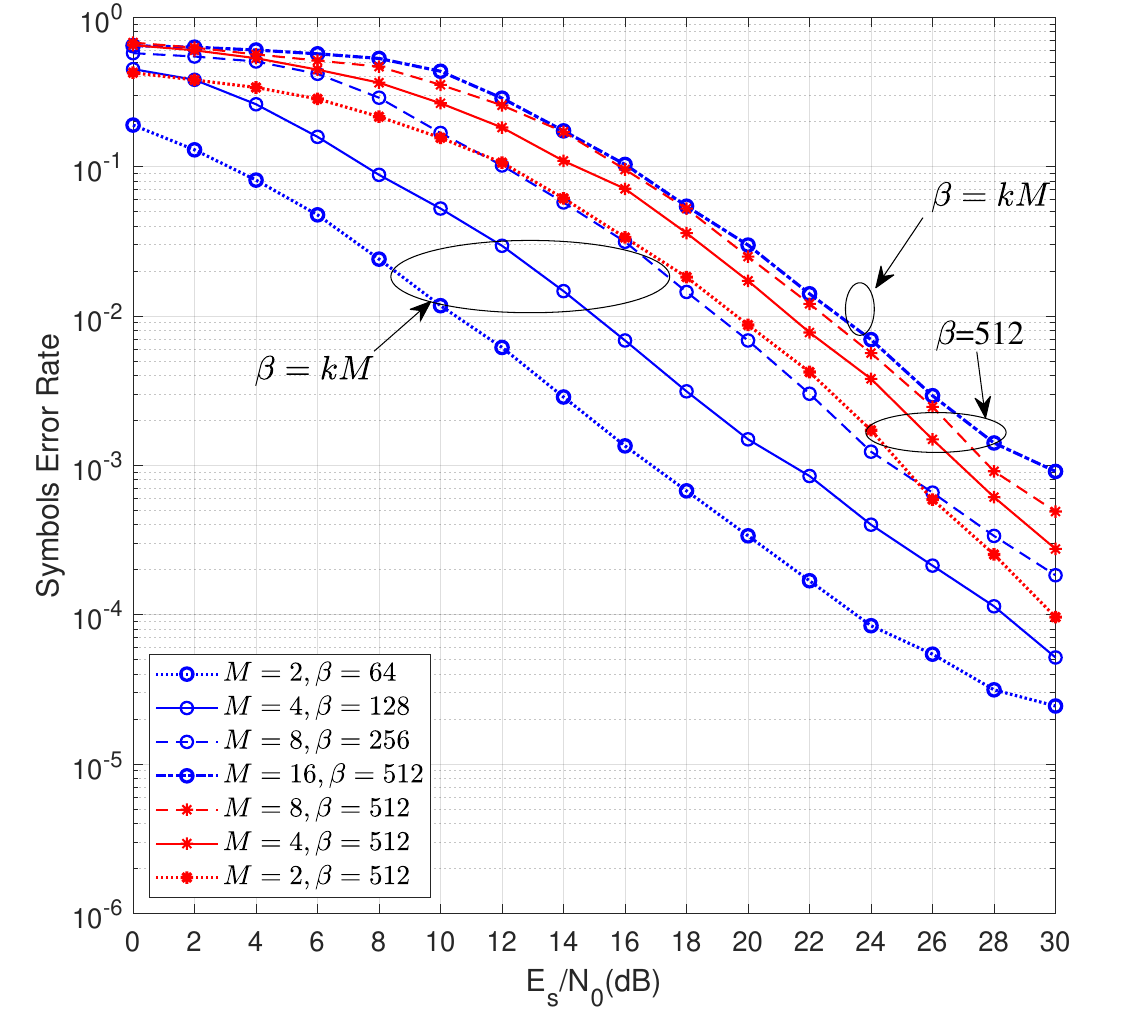}
    \caption{}
\end{subfigure}
\captionsetup{justification=raggedright,singlelinecheck=false}
\caption{SER performances of CCS-$M$-CSK system over different channels for different $M$ and $\beta$. (a) AWGN channel. (b) Multipath Rayleigh channel.}
\label{SER performances of M-ary system over different channels for different M and beta}
\end{figure}

Fig.~\ref{SER performances of M-ary system over different channels for different M and beta} illustrates the SER performance of the proposed CCS-$M$-CSK system under different values of $M$ in both AWGN and two-path Rayleigh channels. It is observed from Fig.~\ref{SER performances of M-ary system over different channels for different M and beta} that when $\beta=512$ and $k=32$, the system's SER performance decreases as $M$ increases. This is primarily because increasing the number of information bits in the $M$-ary constellation reduces the distance between adjacent constellation points, leading to a higher symbol error probability. Additionally, simulations are conducted for $\beta=kM$ with $k=32$. Fig.~\ref{SER performances of M-ary system over different channels for different M and beta} confirms that the trend of decreasing SER performance with increasing $M$ persists. This phenomenon can be explained by the fact that the larger the spreading factor, the more serious the interference by noise, provided that the same information is carried. In conclusion, it is a good alternative to choose $\beta=kM$ with a proper $k$.

\begin{figure}[!t]
\centering
\begin{subfigure}{0.49\columnwidth}    
    \includegraphics[width=\textwidth]{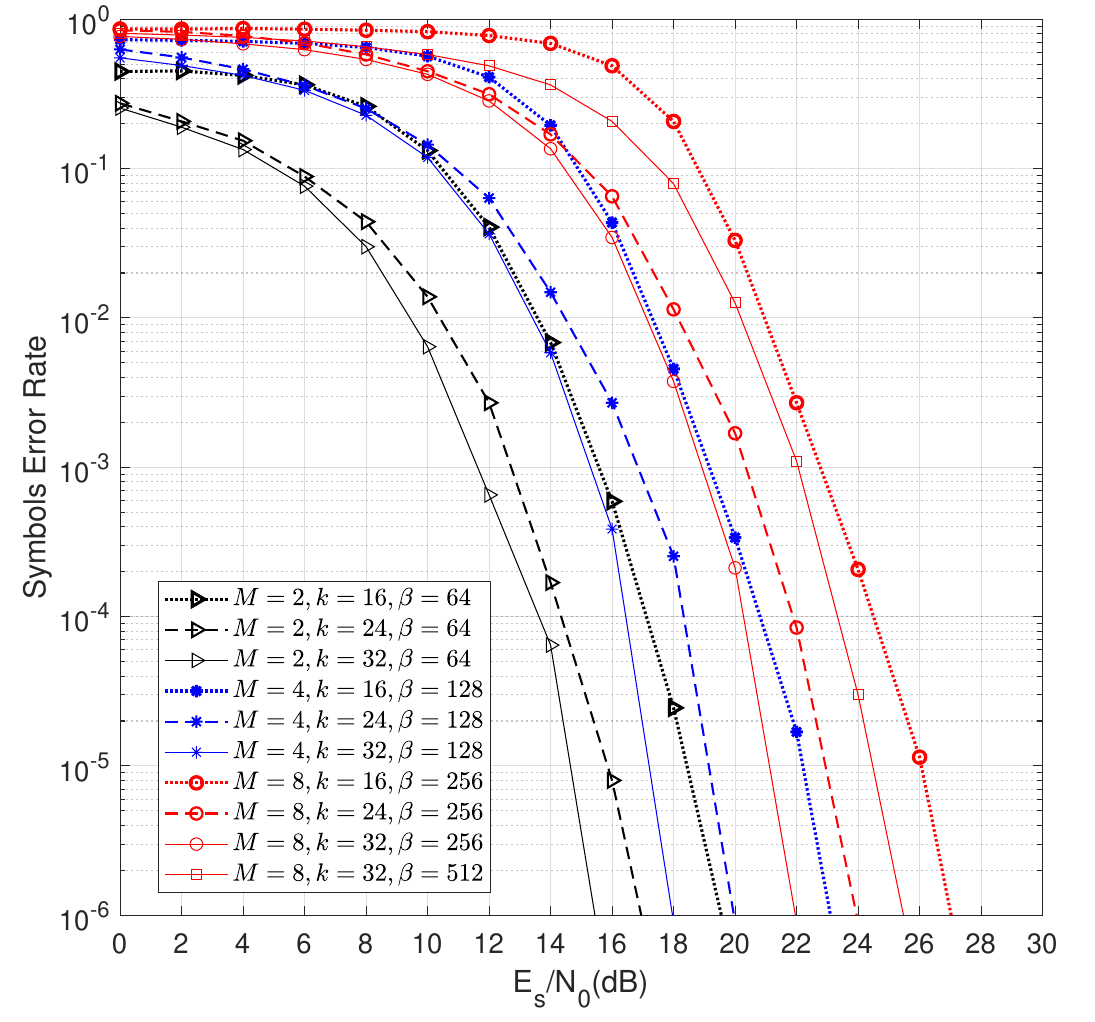}
    \caption{}
\end{subfigure}
\hfill
\begin{subfigure}{0.49\columnwidth}    
    \includegraphics[width=\textwidth]{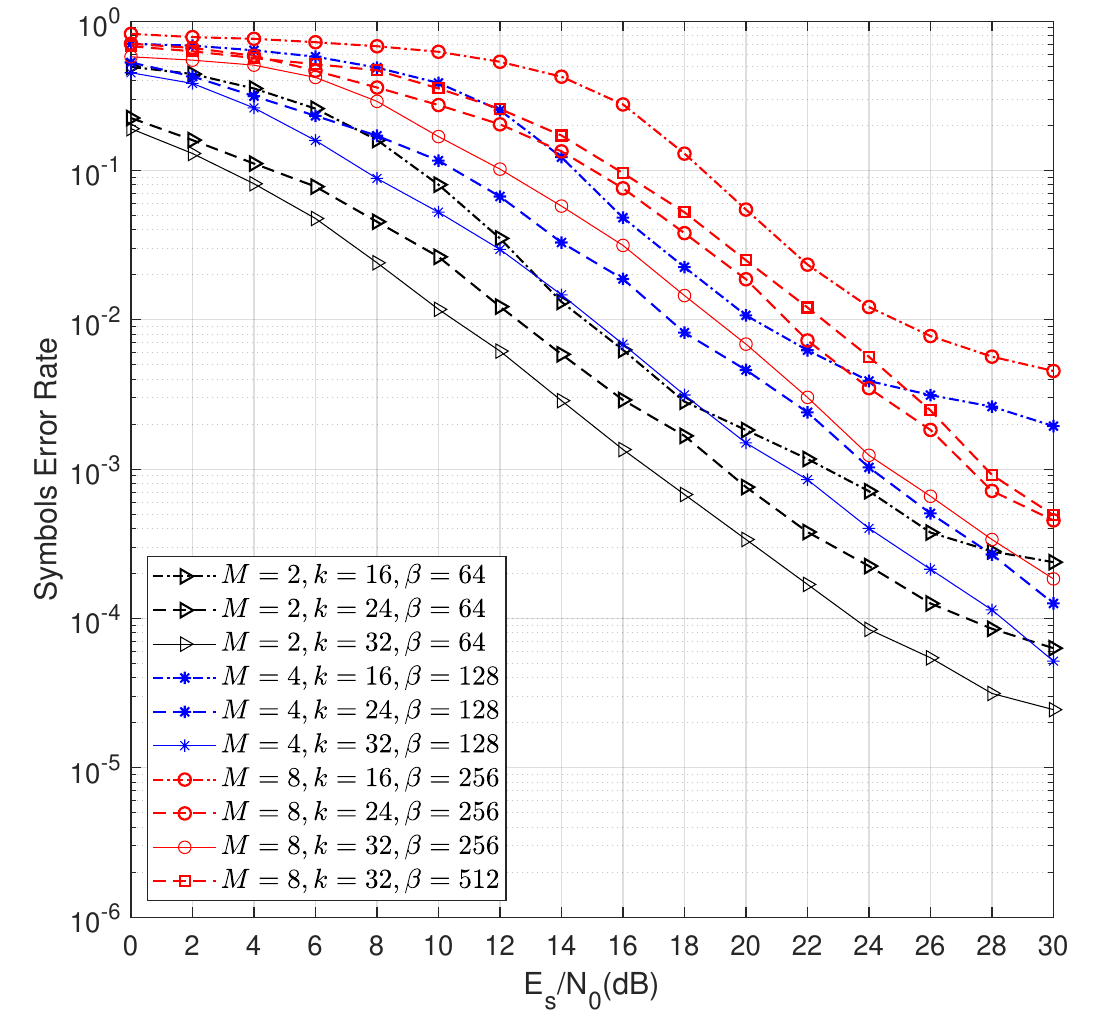}
    \caption{}
\end{subfigure}
\captionsetup{justification=raggedright,singlelinecheck=false}
\caption{SER performances of CCS-$M$-CSK over different channels for different $k$. (a) AWGN channel. (b) Multipath Rayleigh channel.}
\label{SER performances of CCS-$M$-CSK over different channels for different k}
\end{figure}

Fig. \ref{SER performances of CCS-$M$-CSK over different channels for different k} examines how variations in the information-bearing signal $k$ affect the system's SER performance over AWGN and two-path Rayleigh channels. The figures reveal that at a fixed spreading factor, decreasing $k$ leads to a deterioration in the system's SER performance. This is because deep learning-based receivers require sufficiently large $k$ values to effectively extract signal features. Furthermore, with the same $k/\beta$ ratio, i.e., $k=16,\beta=256$ and $k=32,\beta=512$, smaller $k$ values have a more significant impact on SER performance compared to larger $\beta$ values. This indicates that the receiver's ability to accurately extract signal features is more critical than the noise interference on the signal. This phenomenon becomes more evident in Rayleigh multipath channels.

\begin{figure}[!t]
\centering
\begin{subfigure}{0.49\columnwidth}    
    \includegraphics[width=\textwidth]{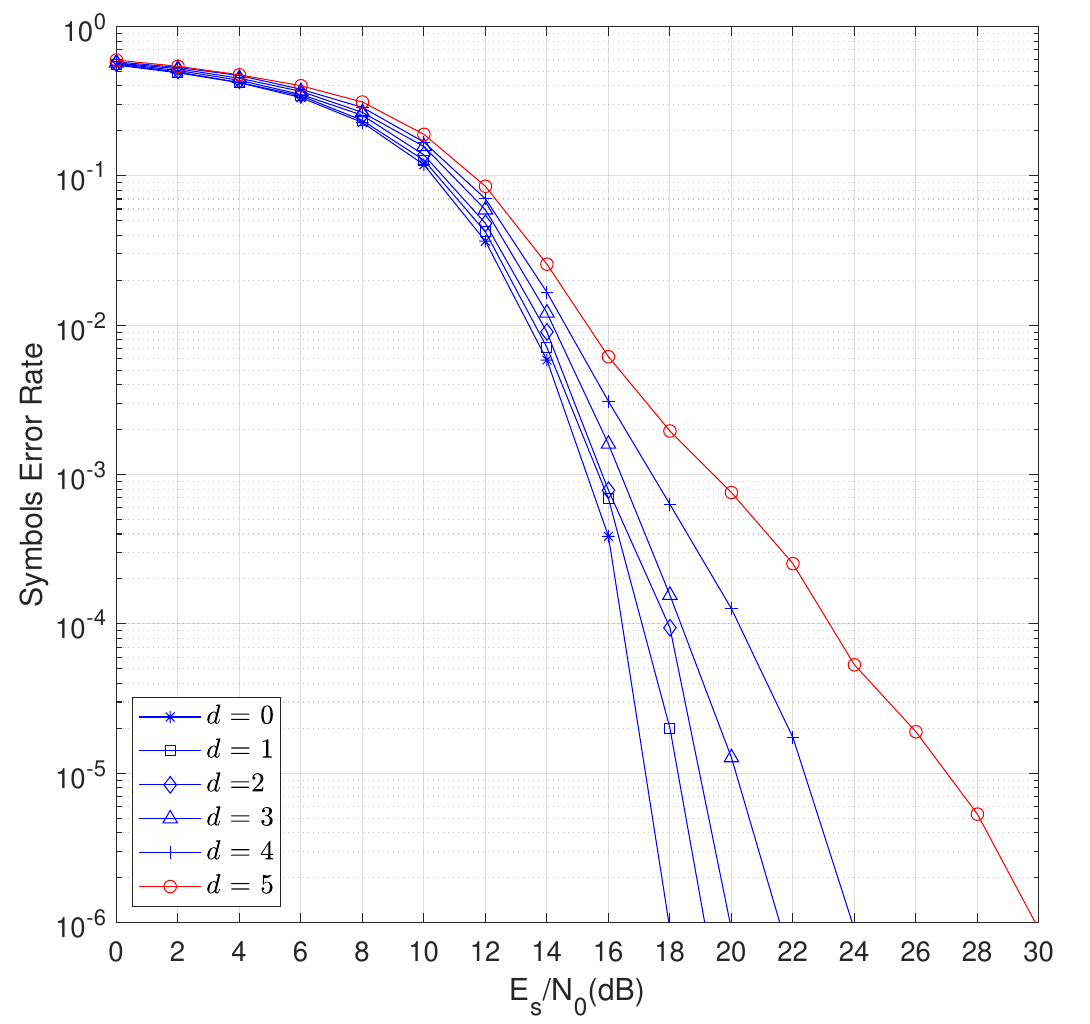}
    \caption{}
\end{subfigure}
\hfill
\begin{subfigure}{0.49\columnwidth}    
    \includegraphics[width=\textwidth]{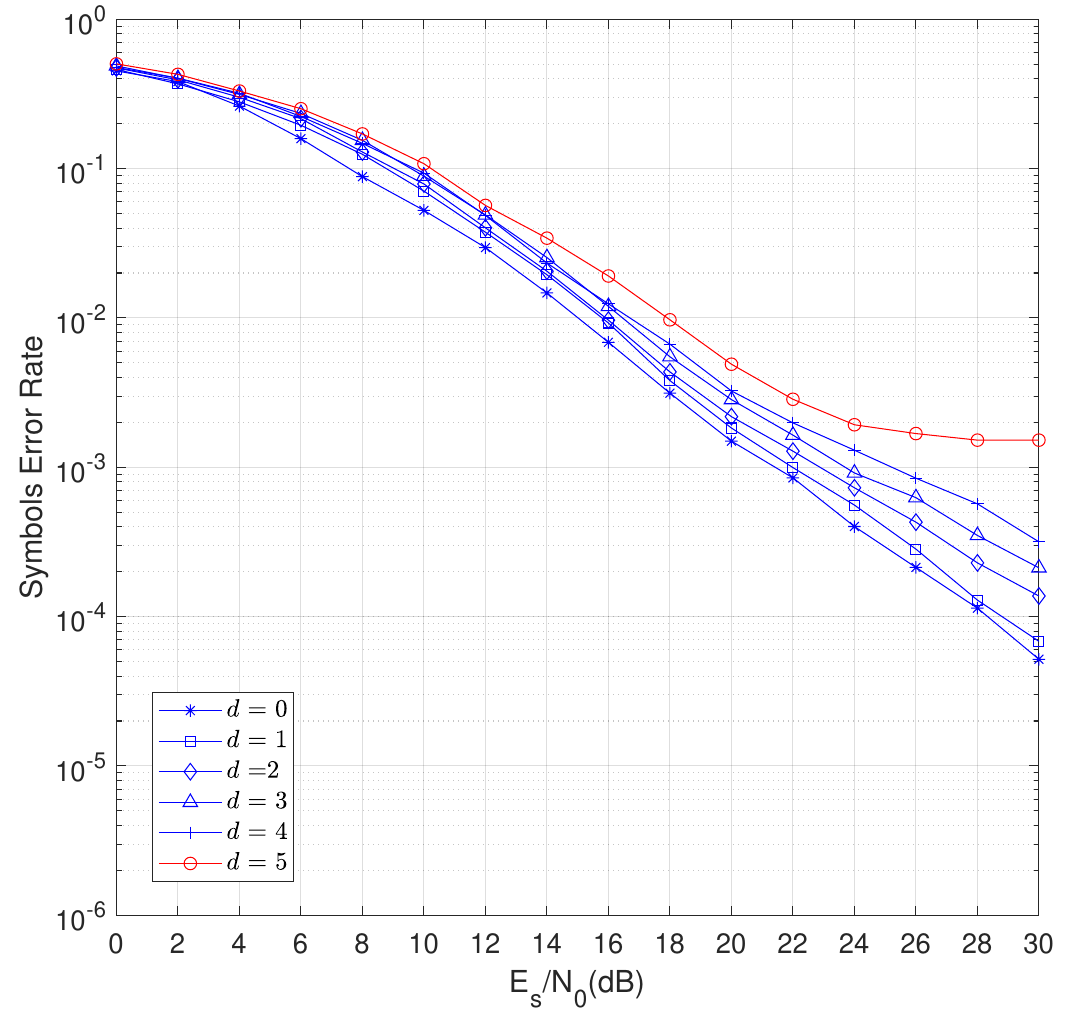}
    \caption{}
\end{subfigure}
\captionsetup{justification=raggedright,singlelinecheck=false}
\caption{The impact of receiving sequence misalignment $d$ on the SER performances over different channels. (a) AWGN channel. (b) Multipath Rayleigh channel.}
\label{The impact of receiving sequence misalignment $d$ on the SER performances over different channels}
\end{figure}

Fig. \ref{The impact of receiving sequence misalignment $d$ on the SER performances over different channels} depicts the impact of receiving sequence misalignment $d$ on the SER performances over AWGN channel and multipath Rayleigh channels, where $k=32$, $M=4$ and $\beta=128$. It is observed from Fig. \ref{The impact of receiving sequence misalignment $d$ on the SER performances over different channels} that whether in AWGN channels or Rayleigh multipath Rayleigh channels, the system performance deteriorates sharply when $d = 5$. Therefore, in order to ensure the system performance, the receiver needs to limit the misalignment length of the received sequence to $d \leq 4$.

\subsection{Comparision of the BER Performance }
\begin{figure}[!t]
\centering
\includegraphics[width=2in]{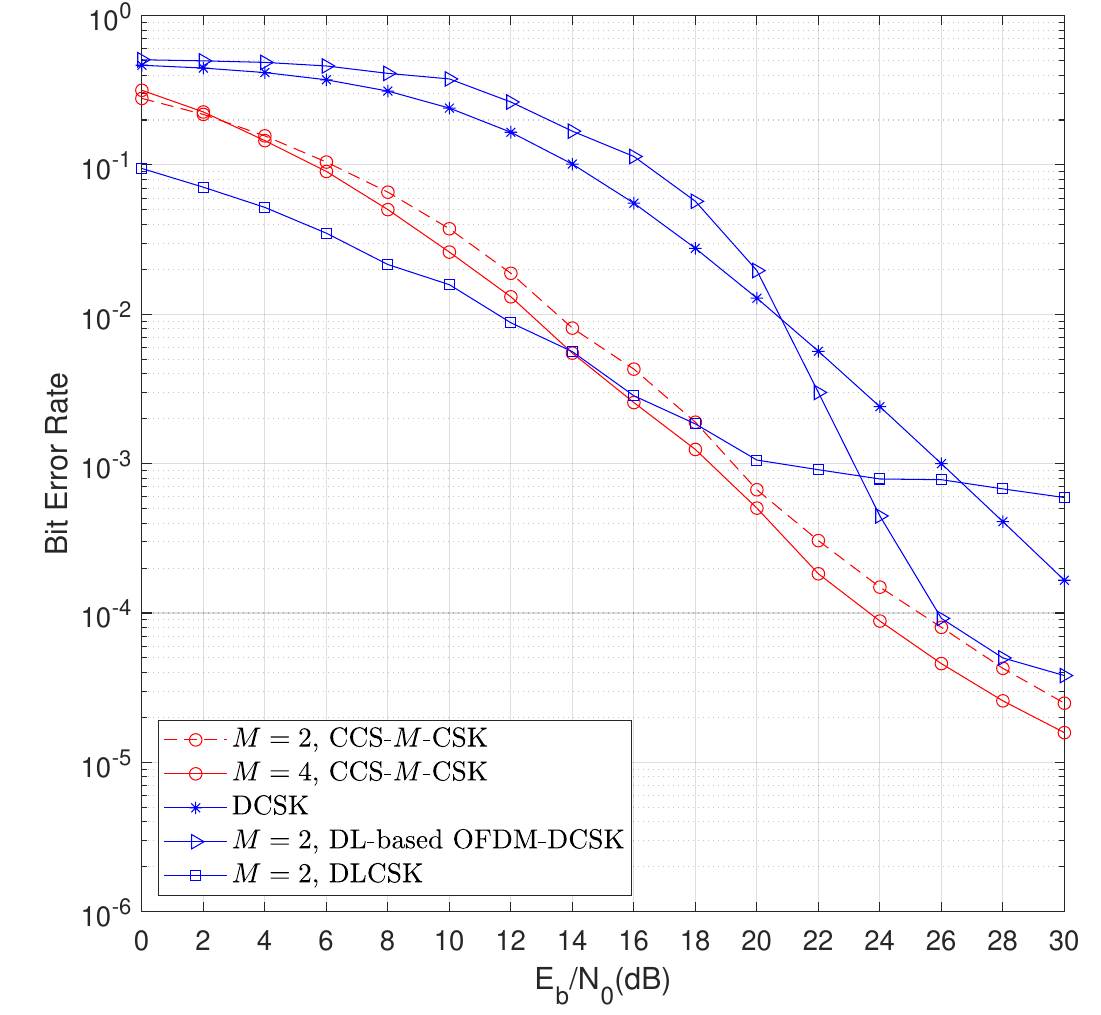}
\captionsetup{justification=raggedright,singlelinecheck=false}
\caption{BER performance comparision of CCS-$M$-CSK, DCSK, DL-based OFDM-DCSK\cite{r3-1} and DLCSK\cite{r3-2}.}
\label{sim5}
\end{figure}

Fig. \ref{sim5} compares BER performances between the proposed system and benchmark systems, i.e., DCSK, OFDM-DCSK and 2-ary DLCSK systems under multipath Rayleigh channels, for $\sigma(n)_{\text{tr}} \in [15, 19]~\text{dB}$, $k=32,\beta=128$. As shown in Fig. \ref{sim5} that the proposed system shows a considerable BER enhancement under multipath Rayleigh channel. Specifically, when $M=2$, $BER=10^{-3}$ , the proposed CCS-$M$-CSK system obtains a performance gain of 1 dB and 4 dB individually compared to DL-based OFDM-DCSK and DLCSK system.

\begin{figure}[!t]
\centering
\begin{subfigure}{0.49\columnwidth}    
    \includegraphics[width=\textwidth,height=1.6in]{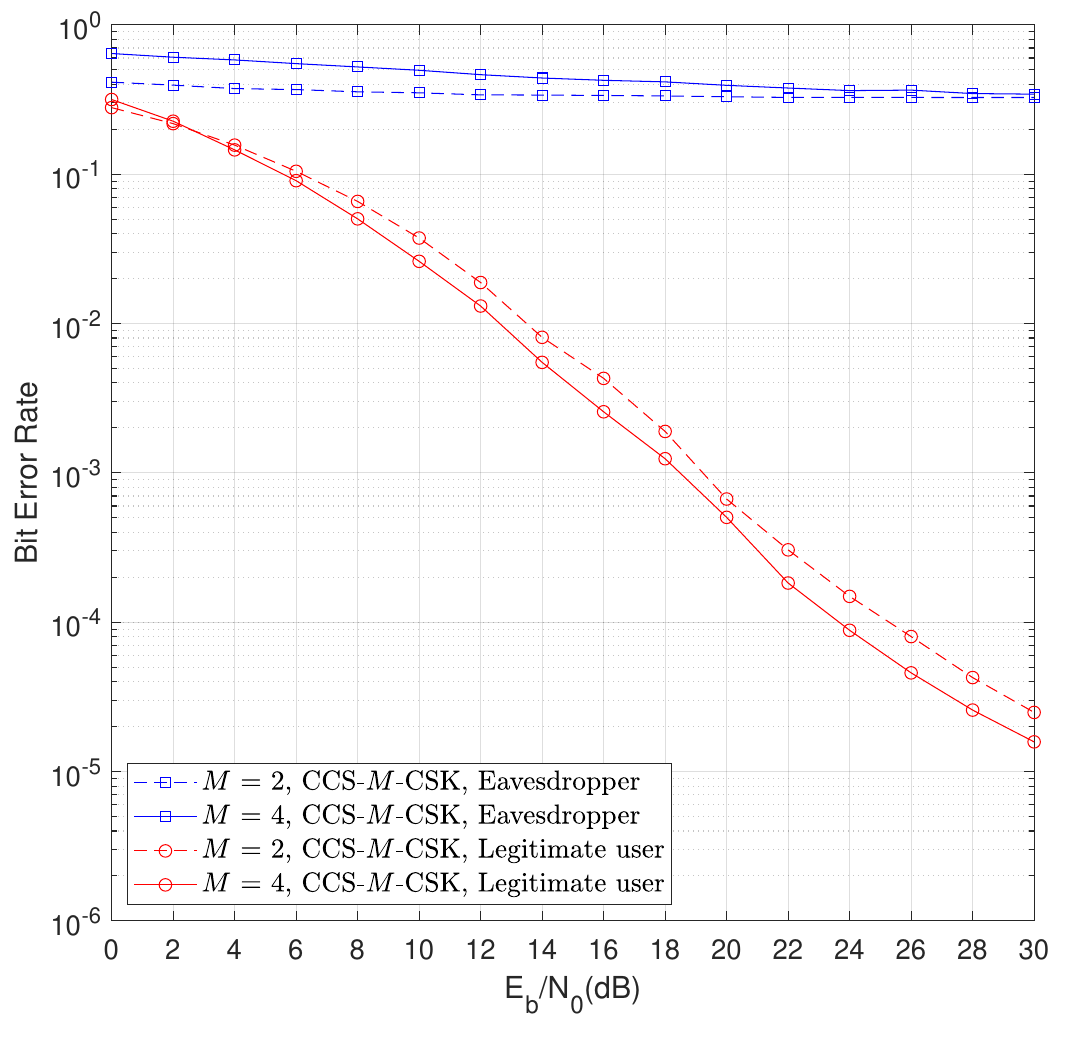}
    \caption{}
\end{subfigure}
\hfill
\begin{subfigure}{0.49\columnwidth}    
    \includegraphics[width=\textwidth,height=1.6in]{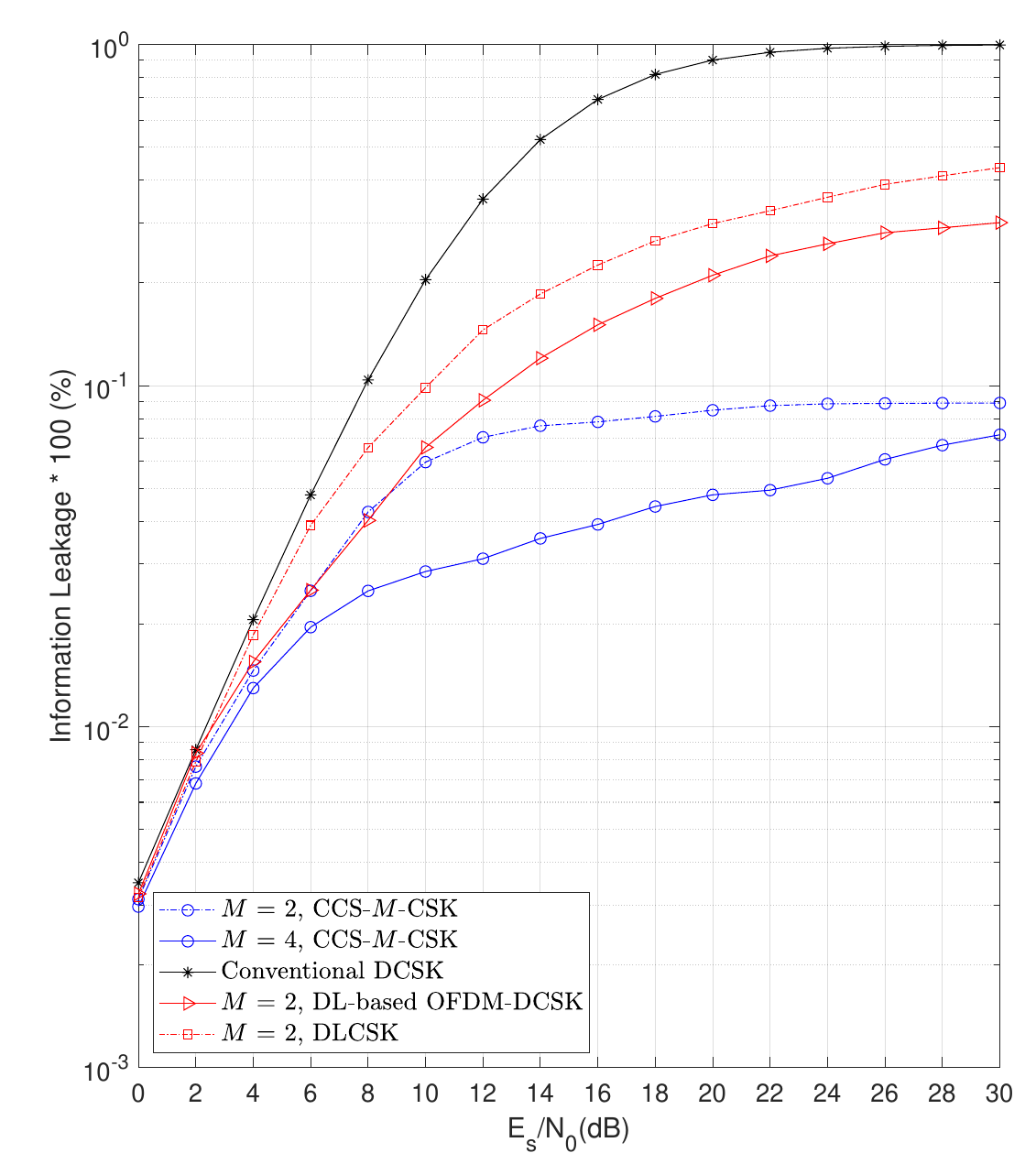}
    \caption{}
\end{subfigure}
\captionsetup{justification=raggedright,singlelinecheck=false}
\caption{Security performance of CCS-$M$-CSK system. (a) BER of eavesdroppers and legitimate user. (b) Information leakage rate of CCS-$M$-CSK system.}
\label{Security performance of CCS-$M$-CSK system}
\end{figure}

Fig. \ref{Security performance of  CCS-$M$-CSK system} investigates the BER and information leakage rate under multipath Rayleigh channel for legitimate users and eavesdroppers with $k=32, \beta=128$. It is assumed that the eavesdroppers are awake of key signaling features, such as symbol duration and symbol onset. For a fair comparison, the eavesdroppers use the same DNN structure but must train it with their captured data.
Fig.\ref{Security performance of  CCS-$M$-CSK system}(a) shows that the legitimate users achieve reliable communication, while the eavesdroppers' performance degrades significantly even when using a matching demodulation technique. This large performance gap arises from the underlying training data asymmetry,i.e, legitimate users can train DNNs using the correct dataset, whereas eavesdroppers must first estimate bits from received signals to construct their training set, leading to error propagation and cumulative performance degradation.
Fig. \ref{Security performance of  CCS-$M$-CSK system}(b) shows that the information leakage rate of the proposed system is consistently low compared to DL-based OFDM-DCSK and DLCSK system.

\section{Conclusion}
In this paper, we have introduced a CCS-$M$-CSK system that simplifies the conventional design by eliminating the need for $M$ chaotic mappings and $M$ classifiers. Instead, our system relies on only two chaotic mappings and binary classification. Compared to the existing CSK-based schemes, our approach offers significant performance improvements. Moreover, this system not only maintains the security benefits inherent in CSK-based designs, but also enhances security based on location. Additionally, when the received sequence is misaligned, our system is capable of maintaining a relatively superior performance for misalignment lengths $d \leq 4$.


\printbibliography
 




\vfill

\end{document}